\documentclass[12pt]{article}
\usepackage{amsfonts}
\usepackage{amssymb}
\usepackage{amsmath}
\usepackage{graphicx}
\usepackage{latexsym}
\usepackage{feynmp}

\newcommand{\dd}{\textrm{d}}
\newcommand{\prt}{\partial}
\newcommand{\strc}{\stackrel{\star}{,}}
\newcommand{\be}{\begin{equation}}
\newcommand{\ee}{\end{equation}}
\newcommand{\bea}{\begin{eqnarray}}
\newcommand{\eea}{\end{eqnarray}}
\newcommand{\pt}{\quad \textrm{.}}
\newcommand{\pc}{\quad \textrm{,}}
\newcommand{\whr}{\quad \textrm{, where} \quad}
\newcommand{\mand}{\quad \textrm{and} \quad}

\newcommand{\mt}{\textrm}
\newcommand{\fsl}{\!\!\!/}

\long\def\symbolfootnote[#1]#2{\begingroup\def\thefootnote{\fnsymbol{footnote}}\footnote[#1]{#2}\endgroup}

\title{On UV/IR Mixing via Seiberg-Witten Map\\for Noncommutative QED}
\author{Matti Raasakka${}^{1,\dagger}$ and Anca Tureanu${}^{1,2}$ \\ \small\textit{${}^{1}$Department
of Physics, University of Helsinki} \\ \small\textit{
${}^{2}$Helsinki
Institute of Physics,}\\ \small\textit{P.O. Box 64, FIN-00014 Helsinki, Finland} \\
\small\texttt{matti.raasakka@aei.mpg.de},
\small\texttt{anca.tureanu@helsinki.fi}}
\date{}
\begin{document}
\maketitle

\abstract We consider quantum electrodynamics in noncommutative
spacetime by deriving a $\theta$-exact Seiberg-Witten map with
fermions in the fundamental representation of the gauge group as an
expansion in the coupling constant. Accordingly, we demonstrate the
persistence of UV/IR mixing in noncommutative QED with charged
fermions via Seiberg-Witten map, extending the results of Schupp and
You \cite{SchuppYou}.\symbolfootnote[0]{\!\!${}^{\dagger}$Present
address: Max Planck Institute for Gravitational Physics (Albert
Einstein Institute),\\ Am M\"uhlenberg 4, D-14476 Golm, Germany.}

\section{Introduction}
The construction of renormalizable quantum field theories in
noncommutative spacetime endowed with canonical coordinate
commutation relations $[\hat{x}^{\mu},\hat{x}^{\nu}] =
i\theta^{\mu\nu}$ is a long-standing problem, the solution of which
is necessary for the calculation of testable predictions for these
theories. They are expected to give hints of the underlying quantum
structure of spacetime, in particular, due to their appearance in
string theory \cite{SeibergWitten} and in semi-classical situations,
where principles of quantum field theory and general relativity are
combined \cite{Doplicher}. Arguably, the most serious obstacle for
the formulation of these noncommutative quantum field theories is
the so-called UV/IR mixing, giving rise to nonrenormalizable
divergencies, which seem to be a generic property of any quantum
field theory in noncommutative spacetime due to the inherent
infinite range of nonlocality induced by the noncommutativity.
Various solutions have been proposed to cure the problem. (See e.g.
\cite{Blaschke} for a review.)

In their seminal paper \cite{SeibergWitten} on the connection
between noncommutative geometry and string theory, Seiberg and
Witten introduced a mapping, which relates gauge field theories in
noncommutative spacetime to ordinary commutative ones, known as the
Seiberg-Witten map. This mapping has virtues, since many aspects of
gauge theories, such as observables and gauge fixing, are more
easily understood and dealt with in the language of ordinary
theories. On the other hand, it also has certain uniqueness
ambiguities explored in \cite{Asakawa,Suo}. Moreover, it does not
seem to affect at all some problems stemming from the
noncommutativity, an example of which is the no-go theorem
\cite{Chaichian1,Chaichian2}, according to which fields can
transform nontrivially under only two different gauge groups
$U_{\star}(N)$.

Therefore it is interesting to study whether the Seiberg-Witten map
affects the problem of UV/IR mixing. Indeed, it has been argued, for
example in \cite{Bichl}\footnote{In particular, in \cite{Bichl} it
was shown that via Seiberg-Witten map the photon self-energy diagram
can be renormalized up to any \emph{finite} order in $\theta$ by
shifting the nonrenormalizable terms up to the next order.}, that
the mixing of UV and IR sectors of noncommutative theories is absent
in the Seiberg-Witten formalism. However, we suspect that this may
be due to the expansion in the noncommutativity parameter matrix
$\theta$ in the $\theta$-expanded Seiberg-Witten map. In the
$\theta$-exact Seiberg-Witten map for noncommutative QED, the UV/IR
mixing problem does appear, as we shall demonstrate. The same
argument has been expressed by Schupp and You in \cite{SchuppYou},
where they considered a noncommutative model with a gauge field
coupled with a spinor field in the adjoint representation of the
gauge group, and showed the existence of an IR-divergent term for
the photon self-energy corrections. The adjoint representation,
however, corresponds to a chargeless particle but with an electric
dipole moment proportional to $\theta$
\cite{Jabbari,Susskind,Yin,Alvarez}, and thus in their model the
interaction vanishes at the commutative limit $\theta \rightarrow
0$. Therefore the model does not correspond to a noncommutative
theory of electrically charged fermions, which should reduce to the
ordinary QED in the commutative limit.

In this paper, our primary goal is to extend the analysis of Ref.
\cite{SchuppYou} to the case of noncommutative QED with charged
fermions. We first derive a $\theta$-exact Seiberg-Witten map for a
gauge theory with a spinor field in the fundamental representation
of the gauge field, corresponding to charged fermions, as an
expansion in the coupling constant, and then demonstrate the
persistence of UV/IR mixing in the photon self-energy corrections.

\section{$\theta$-exact Seiberg-Witten map with charged fermions}
The Seiberg-Witten map, as introduced in Ref. \cite{SeibergWitten},
is a technique to induce a gauge orbit preserving mapping
$(A_{\mu},\Lambda) \mapsto (\hat{A}_{\mu},\hat{\Lambda})$ between
gauge fields and gauge transformation parameters in commutative and
noncommutative spacetimes, respectively. The mapping has previously
been realized either as an expansion in the noncommutativity
parameters $\theta^{\mu\nu}$ or in the gauge field $A_{\mu}$ as
established for an Abelian gauge field theory in
\cite{Garousi,MehenWise,SchuppYou} (in the respective order).
However, there is a third way, namely, an expansion in the coupling
constant, which is the one we shall use in the
following.\footnote{Since an expansion in $\theta^{\mu\nu}$ may
obscure the possible UV/IR mixing of the noncommutative theory, a
$\theta$-exact approach is essential.} This is particularly
convenient, since for the usual perturbation theory of QED we shall
perform an expansion in the coupling constant in any case.
Accordingly, we are able to avoid performing multiple expansions by
taking terms of the expansion series of Seiberg-Witten map
appropriately into account. We will also add a spinor field in the
fundamental representation of the gauge field into the picture, thus
inducing a mapping $(\Psi,A_{\mu},\Lambda) \mapsto
(\hat{\Psi},\hat{A}_{\mu},\hat{\Lambda})$.

The strategy in deriving the $\theta$-exact Seiberg-Witten map, in a
nutshell, is first to relate two gauge field theories in
noncommutative spacetimes with infinitesimally differing
noncommutativity parameter matrices, say $\theta$ and $\theta'$, to
each other in a gauge orbit preserving way, and then to integrate
this relation from the origin $\theta_{0} \equiv 0$ to some constant
matrix $\theta_{1}$ along a path in the space of $4 \times 4$
real-valued anti-symmetric matrices. Thus, let us have two
noncommutative gauge field theories with spinor fields, denoted by
$\mathcal{T}[\theta^{\mu \nu},A_{\mu},\Psi]$ and
$\mathcal{T}'[\theta'^{\mu \nu},A'_{\mu},\Psi']$, where the
arguments are the noncommutativity parameters, the gauge fields and
the spinor fields, respectively. Let us also introduce the notation
\bea
    \theta'^{\mu \nu} - \theta^{\mu \nu} &=& \delta\theta^{\mu \nu} \pc \nonumber\\
    A'_{\mu} - A_{\mu} &=& a_{\mu} \pc\nonumber\\
    \Psi' - \Psi &=& \psi \pt \label{eq:fielddiff}
\eea As prescribed, we assume that $\delta\theta^{\mu \nu}$ are
infinitesimal, and that the fields depend smoothly on the
noncommutativity parameters, so that $a_{\mu}$, $\psi$ and all their
partial derivatives are also infinitesimal.

Let us now consider a mapping of the fields from $\mathcal{T}$ to
$\mathcal{T}'$. We may think of the fields in $\mathcal{T}'$ as
depending on the fields in $\mathcal{T}$ according to this mapping,
namely\footnote{Precisely which arguments are needed here depends
on, and is revealed by, the solutions found below, but for clarity
they are already given here. Moreover, we have dropped the Lorentz
indices of the arguments for simplicity, since it is clear how they
are resumed.}
\be
 A'_{\mu} \equiv A'_{\mu}(A) = A_{\mu} + a_{\mu}(A) \quad \textrm{and} \quad \Psi' \equiv \Psi'(\Psi,A) = \Psi + \psi(\Psi,A) \pt
\ee
Now, we apply a gauge transformation in the theory $\mathcal{T}$
with a gauge transformation parameter $\Lambda$. For a
noncommutative gauge field theory a gauge transformation is given by
the formulae\footnote{We do not worry about gauge fixing here, since
we shall ultimately fix it in the commutative QED to which we arrive
in Section \ref{sec:ncqed}. Therefore, the Faddeev-Popov ghost
fields are not needed.} \bea
    \delta_{\Lambda} A_{\mu} &=& \prt_{\mu} \Lambda + ig[\Lambda \stackrel{\star}{,}A_{\mu}] \pc \nonumber\\
    \delta_{\Lambda} \Psi &=& i g \Lambda \star \Psi \pc \label{eq:gaugetransf}
\eea where $g$ is the coupling constant and the noncommutative
$\star$-product is the Moyal product defined as \be
    f \star g = \left. \exp \left[ \frac{i}{2}\theta^{\mu\nu} \frac{\prt}{\prt x^{\mu}} \frac{\prt}{\prt y^{\nu}} \right] f(x) g(y) \right|_{x=y} \pt
\ee The fundamental requirement for the Seiberg-Witten map is that
it should preserve the gauge equivalence classes of the theory, so
that the transformation $\Lambda$ in $\mathcal{T}$ corresponds to a
gauge transformation \be
    \Lambda' \equiv \Lambda'(\Lambda,A) = \Lambda + \lambda(\Lambda,A)
\ee in $\mathcal{T}'$: \bea
    A'_{\mu}(A + \delta_{\Lambda} A) &=& A'_{\mu}(A) + \delta_{\Lambda'} A'_{\mu}(A) \pc\label{eq:SWreq1}\\
    \Psi'(\Psi + \delta_{\Lambda} \Psi , A + \delta_{\Lambda} A) &=& \Psi'(\Psi,A) + \delta_{\Lambda'} \Psi'(\Psi,A) \pt \label{eq:SWreq}
\eea By substituting the formulae (\ref{eq:fielddiff}) and
(\ref{eq:gaugetransf}) into \eqref{eq:SWreq1} and \eqref{eq:SWreq},
and using the relation \be
    f \star' g = f\ e^{\frac{i}{2} \overleftarrow{\prt_{\mu}}
(\theta + \delta\theta)^{\mu\nu} \overrightarrow{\prt_{\nu}}}\ g = f
\star g + \frac{i}{2} \delta\theta^{\mu\nu} (\prt_{\mu}f) \star
(\prt_{\nu}g) \pc \ee we arrive at the equations \bea
    && a_{\mu}(A + \delta_{\Lambda}A) - a_{\mu}(A) - \prt_{\mu}\lambda(\Lambda,A) - ig\left[ \lambda(\Lambda,A) \strc A_{\mu} \right] - ig\left[ \Lambda \strc a_{\mu}(A) \right] \nonumber\\
    &=& -\frac{g}{2} \delta\theta^{\alpha\beta} \left\{ \prt_{\alpha}\Lambda \strc \prt_{\beta}A_{\mu} \right\} \label{eq:Adiff}
\eea and \bea
    && \psi(\Psi + \delta_{\Lambda}\Psi,A + \delta_{\Lambda}A) - \psi(\Psi,A) - ig\Lambda \star \psi(\Psi,A) - ig\lambda(\Lambda,A) \star \Psi \nonumber\\
    &=& -\frac{g}{2} \delta\theta^{\alpha\beta} (\prt_{\alpha}\Lambda) \star (\prt_{\beta}\Psi) \label{eq:Psidiff}
\eea for $\lambda$, $a_{\mu}$ and $\psi$. As found by Seiberg and
Witten in \cite{SeibergWitten} (for $g \equiv 1$), the equation
(\ref{eq:Adiff}) is solved by \bea
    \lambda &=& -\frac{g}{4} \delta\theta^{\alpha\beta} \left\{ A_{\alpha} \strc \prt_{\beta}\Lambda \right\} \pc \nonumber\\
    a_{\mu} &=& -\frac{g}{4} \delta\theta^{\alpha\beta} \left\{ A_{\alpha} \strc \prt_{\beta}A_{\mu} + F_{\beta\mu} \right\} \pc \label{eq:gaugesol}
\eea where $F_{\mu\nu} \equiv \prt_{\mu}A_{\nu} - \prt_{\nu}A_{\mu}
- ig\left[ A_{\mu} \strc A_{\nu} \right]$ is the field strength.
Using (\ref{eq:gaugesol}), we find for the equation
(\ref{eq:Psidiff}) the solution \be\label{eq:psisol}
    \psi = -\frac{g}{2} \delta\theta^{\alpha\beta} \left[ A_{\alpha} \star (\prt_{\beta}\Psi) + \frac{1}{2} (\prt_{\beta}A_{\alpha}) \star \Psi \right] \pt
\ee

As prescribed, the next step in constructing the $\theta$-exact
Seiberg-Witten map is to integrate these relations along a path in
the space of real-valued anti-symmetric matrices to obtain a
relation between gauge theories in a commutative spacetime and in a
noncommutative one with finite noncommutativity parameters
$\theta^{\mu\nu}$. There are certain ambiguities related to choosing
a particular path, following from the observation that successive
Seiberg-Witten maps do not commute in general, and thus there is an
infinite number of free parameters related to the path fixing. Some
but not all of these correspond to gauge transformations and field
redefinitions, as explored in \cite{Asakawa,Suo}. However, for
simplicity, we choose to consider a straight path\footnote{This is
the case considered also  in Ref. \cite{SchuppYou}.} $\gamma: [0,1]
\rightarrow \{\theta \in \mathbb{R}^{4 \times 4} | \theta\
\textrm{antisymmetric}\}$ such that $\gamma(s) = s\theta_{1}$, where
$\theta_{1}$ is the constant matrix reached at $s=1$. Let us denote
the fields, now considered as dependent on the spacetime coordinates
$x^{\mu}$ and the noncommutativity parameters $\theta^{\mu\nu}$, as
$A_{\mu}(x;\theta)$ and $\Psi(x;\theta)$. Integrating the variation
(\ref{eq:Adiff}) along the straight path $\gamma$ and using
integration by parts, we get for the gauge field the equation \bea
    A_{\mu}(x;\theta_{1}) &=& A_{\mu}(x;0) + \lim_{y \rightarrow x} \Bigg\{ -\frac{g\theta_{1}^{\alpha\beta}}{4} \frac{e^{\frac{i}{2}\theta^{\rho\sigma} \frac{\prt}{\prt x^{\rho}} \frac{\prt}{\prt y^{\sigma}}}}{ \frac{i}{2}\theta_{1}^{\gamma\delta} \frac{\prt}{\prt x^{\gamma}} \frac{\prt}{\prt y^{\delta}}} \nonumber\\
    && \quad \times \Big[ A_{\alpha}(x;\theta) \big( \prt_{\beta} A_{\mu}(y;\theta) + F_{\beta\mu}(y;\theta) \big) \nonumber\\
    && \qquad + \big( \prt_{\beta} A_{\mu}(x;\theta) + F_{\beta\mu}(x;\theta) \big) A_{\alpha}(y;\theta) \Big] \nonumber\\
    && + \frac{g\theta_{1}^{\alpha\beta}}{4} \sum_{n=2}^{\infty} (-1)^{n} \frac{e^{\frac{i}{2}\theta^{\rho\sigma} \frac{\prt}{\prt x^{\rho}} \frac{\prt}{\prt y^{\sigma}}}}{ \left( \frac{i}{2}\theta_{1}^{\gamma\delta} \frac{\prt}{\prt x^{\gamma}} \frac{\prt}{\prt y^{\delta}} \right)^{n} } \left( \prod_{k=2}^{n} \theta_{1}^{\alpha_{k}\beta_{k}} \frac{\delta}{\delta\theta^{\alpha_{k}\beta_{k}}} \right) \nonumber\\
    && \quad \times \Big[ A_{\alpha}(x;\theta) \big( \prt_{\beta} A_{\mu}(y;\theta) + F_{\beta\mu}(y;\theta) \big) \nonumber\\
    && \qquad + {\big( \prt_{\beta} A_{\mu}(x;\theta) + F_{\beta\mu}(x;\theta) \big) A_{\alpha}(y;\theta) \Big] \Bigg\}}_{\theta=0}^{\theta=\theta_{1}} \mt{,} \label{eq:Aiter}
\eea and similarly for the spinor field the equation \bea
    \Psi(x;\theta_{1}) &=& \Psi(x;0) + \lim_{y \rightarrow x} \Bigg\{ -\frac{g\theta_{1}^{\alpha\beta}}{4} \frac{e^{\frac{i}{2}\theta^{\rho\sigma} \frac{\prt}{\prt x^{\rho}} \frac{\prt}{\prt y^{\sigma}}}}{ \frac{i}{2}\theta_{1}^{\gamma\delta} \frac{\prt}{\prt x^{\gamma}} \frac{\prt}{\prt y^{\delta}}} \nonumber\\
    && \quad \times \Big[ A_{\alpha}(x;\theta) (\prt_{\beta} \Psi(y;\theta)) + \frac{1}{2} (\prt_{\beta} A_{\alpha}(x;\theta)) \Psi(y;\theta) \Big] \nonumber\\
    && + \frac{g\theta_{1}^{\alpha\beta}}{4} \sum_{n=2}^{\infty} (-1)^{n} \frac{e^{\frac{i}{2}\theta^{\rho\sigma} \frac{\prt}{\prt x^{\rho}} \frac{\prt}{\prt y^{\sigma}}}}{ \left( \frac{i}{2}\theta_{1}^{\gamma\delta} \frac{\prt}{\prt x^{\gamma}} \frac{\prt}{\prt y^{\delta}} \right)^{n} } \left( \prod_{k=2}^{n} \theta_{1}^{\alpha_{k}\beta_{k}} \frac{\delta}{\delta\theta^{\alpha_{k}\beta_{k}}} \right) \nonumber\\
    && \quad \times {\Big[ A_{\alpha}(x;\theta) (\prt_{\beta} \Psi(y;\theta)) + \frac{1}{2} (\prt_{\beta} A_{\alpha}(x;\theta)) \Psi(y;\theta) \Big] \Bigg\}}_{\theta=0}^{\theta=\theta_{1}} \mt{,}\nonumber\\ \label{eq:Psiiter}
\eea which can be calculated iteratively in powers of the coupling
constant $g$, since $\frac{\delta}{\delta\theta}A_{\mu} =
\mathcal{O}(g)$ and $\frac{\delta}{\delta\theta}\Psi =
\mathcal{O}(g)$, so the variations in the sums give terms of ever
increasing powers in $g$.

\section{NCQED via Seiberg-Witten map}\label{sec:ncqed}
\begin{figure}
    \centering\includegraphics[width=0.6\linewidth]{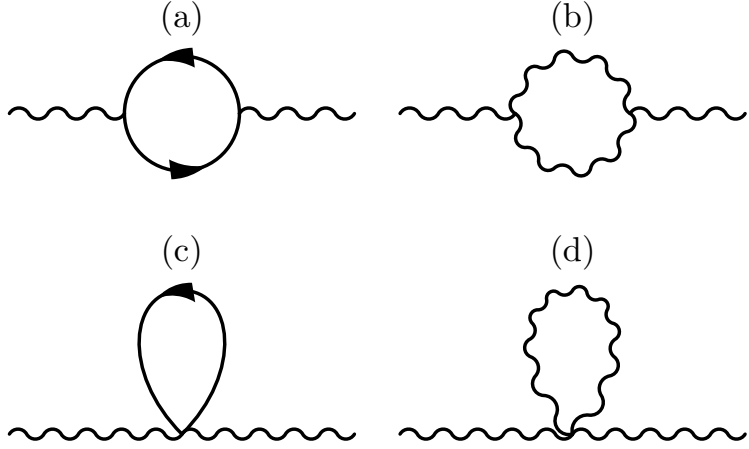}
    \caption[selfenergy]{\small \label{fig:selfenergy} Photon self-energy diagrams in the second order of $e$.}
\end{figure}

We now turn to consider exclusively the gauge group $U(1)$, i.e.,
quantum electrodynamics (QED). We need to study the action of
noncommutative QED, \be\label{eq:ncqedaction}
    \mathcal{S}_{\textrm{\tiny NCQED}} = \int \dd^{4}x
    \left[ \hat{\bar{\Psi}} \star (i \prt\fsl - m) \hat{\Psi} -\frac{1}{4} \hat{F}_{\mu\nu} \star \hat{F}^{\mu\nu} -
    e\hat{\bar{\Psi}} \star \hat{A}\fsl \star \hat{\Psi} \right]\,,
\ee in terms of the ordinary fields up to the second order in the
electromagnetic coupling constant $e$ in order to catch all the
second order contributions to the photon self-energy. These arise
from the diagrams drawn in Fig.  \ref{fig:selfenergy}.
Denoting the noncommutative fields by hats and dropping the lower
index from $\theta_{1}$, we find the gauge field via the equation
(\ref{eq:Aiter}) up to second order in the coupling constant:
 \bea
\label{eq:hatA}
    \hat{A}_{\mu}^{(0)} &=& A_{\mu} \pc \nonumber\\
    \hat{A}_{\mu}^{(1)} &=& -e \frac{\sin(\frac{1}{2}\prt_{1} \wedge \prt_{2})}{\frac{1}{2}\prt_{1} \wedge \prt_{2}} \Big[ \eta_{\mu}^{\beta} \tilde{\prt}_{2}^{\alpha} - \frac{1}{2} \theta^{\alpha\beta}\prt_{2\mu} \Big] A_{\alpha}(x_{1}) A_{\beta}(x_{2}) \Bigg|_{x_{1}=x_{2}} \nonumber\\
    &=& -\frac{e}{2} \theta^{\alpha\beta} A_{\alpha} \star_{1}^{s} (2\prt_{\beta} A_{\mu} - \prt_{\mu}A_{\beta}) \pc\nonumber\\
    \hat{A}_{\mu}^{(2)} &=& \frac{e^{2}}{4} \Bigg[ \frac{\sin(\frac{1}{2} \prt_{1} \wedge \prt_{2})}{\frac{1}{2} \prt_{1} \wedge \prt_{2}} \frac{\sin(\frac{1}{2}(\prt_{1} + \prt_{2}) \wedge \prt_{3})}{\frac{1}{2}(\prt_{1} + \prt_{2}) \wedge \prt_{3}} \nonumber\\
    && \qquad \qquad + \frac{ \cos(\frac{1}{2} \prt_{1} \wedge \prt_{2}) \cos(\frac{1}{2} (\prt_{1} + \prt_{2}) \wedge \prt_{3}) - 1}{ [ \frac{1}{2}(\prt_{1} + \prt_{2}) \wedge \prt_{3} ]^{2}} \Bigg] \nonumber\\
    && \times \Big\{ 2[2\tilde{\prt}_{2}^{\alpha}\tilde{\prt}_{3}^{\beta} - \theta^{\alpha\beta} \prt_{2} \wedge \prt_{3}] \eta_{\mu}^{\rho} + 2[2(\tilde{\prt}_{1} + \tilde{\prt}_{2})^{\rho}\tilde{\prt}_{2}^{\alpha} - \theta^{\rho\alpha}\prt_{1}\wedge\prt_{2}]\eta_{\mu}^{\beta} \nonumber\\
    && \qquad + [(2\theta^{\beta\rho}\tilde{\prt}_{2}^{\alpha} - \theta^{\alpha\beta}(3\tilde{\prt}_{1} + \tilde{\prt}_{2})^{\rho} - 2\theta^{\rho\alpha}\tilde{\prt}_{1}^{\beta})\prt_{2\mu} \nonumber\\
    && \qquad \quad - (\theta^{\alpha\beta}\tilde{\prt}_{2}^{\rho} + 2\theta^{\beta\rho}\tilde{\prt}_{2}^{\alpha})\prt_{3\mu}]
    \Big\} A_{\alpha}(x_{1}) A_{\beta}(x_{2}) A_{\rho}(x_{3}) \Bigg|_{x_{1}=x_{2}=x_{3}}\,. \eea
    Similarly for the spinor field, via the equation
(\ref{eq:Psiiter}), we find: \bea \label{eq:hatPsi}
    \hat{\Psi}^{(0)} &=& \Psi \pc \nonumber \\
    \hat{\Psi}^{(1)} &=& -\frac{e}{2} \frac{e^{\frac{i}{2} \prt_{1} \wedge \prt_{2}} - 1}{\frac{i}{2} \prt_{1} \wedge \prt_{2}} (\frac{1}{2}\tilde{\prt}_{1} + \tilde{\prt}_{2})^{\alpha} A_{\alpha}(x_{1}) \Psi(x_{2}) \Bigg|_{x_{1}=x_{2}} \nonumber\\
    &=& -\frac{e}{2} \theta^{\alpha\beta} \left[ A_{\alpha} \star_{1} (\prt_{\beta}\Psi) + \frac{1}{2}(\prt_{\beta}A_{\alpha}) \star_{1} \Psi \right] \pc \nonumber\\
    \hat{\Psi}^{(2)} &=& \frac{e^{2}}{4} \Bigg\{ \Bigg[ \frac{ \sin( \frac{1}{2} \prt_{1} \wedge \prt_{2}  )}{\frac{1}{2}\prt_{1} \wedge \prt_{2} } \frac{ e^{ \frac{i}{2}(\prt_{1} + \prt_{2}) \wedge \prt_{3} } }{ \frac{i}{2}(\prt_{1} + \prt_{2}) \wedge \prt_{3} } - \frac{ \cos( \frac{1}{2} \prt_{1} \wedge \prt_{2} ) e^{\frac{i}{2} (\prt_{1} + \prt_{2}) \wedge \prt_{3}} + 1 }{ [ \frac{1}{2} (\prt_{1} + \prt_{2}) \wedge \prt_{3} ]^{2}} \Bigg] \nonumber\\
    && \qquad \times \Big[ 2\tilde{\prt}_{2}^{\alpha} (\frac{1}{2}\tilde{\prt}_{1} + \frac{1}{2}\tilde{\prt}_{2} + \tilde{\prt}_{3})^{\beta} + \theta^{\alpha\beta} (\frac{1}{2}\prt_{1} + \prt_{3}) \wedge \prt_{2} \Big] \nonumber\\
    && \quad + \Bigg[ \frac{ e^{\frac{i}{2} \prt_{1} \wedge \prt_{3}} - 1}{\frac{i}{2} \prt_{1} \wedge \prt_{3}} \frac{ e^{\frac{i}{2} \prt_{2} \wedge (\prt_{1} + \prt_{3})}}{\frac{i}{2} \prt_{2} \wedge (\prt_{1} + \prt_{3})} - \frac{ e^{\frac{i}{2} \prt_{1} \wedge \prt_{3}} e^{\frac{i}{2} \prt_{2} \wedge (\prt_{1} + \prt_{3})} + 1}{ [ \frac{i}{2} \prt_{2} \wedge (\prt_{1} + \prt_{3}) ]^{2}} \Bigg] \nonumber\\
    && \qquad \times (\frac{1}{2}\tilde{\prt}_{1} + \tilde{\prt}_{3})^{\alpha} (\tilde{\prt}_{1} + \frac{1}{2} \tilde{\prt}_{2} + \tilde{\prt}_{3})^{\beta} \Bigg\} A_{\alpha}(x_{1}) A_{\beta}(x_{2}) \Psi(x_{3}) \Bigg|_{x_{1}=x_{2}=x_{3}} \ \mt{,}
\eea where we have introduced the notations $\tilde{u}^{\mu} :=
\theta^{\mu\nu} u_{\nu}$ and $u \wedge v := u_{\mu} \theta^{\mu\nu}
v_{\nu}$ for any four-vectors $u_{\mu}$, $v_{\mu}$,
and\footnote{Notice that our notation for these so-called
`generalized $\star$-products' differs from that used in
\cite{Garousi,MehenWise,SchuppYou}. This is an attempt to make the
notation more systematic. The lower index denotes the times of
integration of the $\star$-product over the unit interval, and the
upper index `$s$' denotes symmetrization of the product with respect
to its arguments.} \bea
    f \star_{1} g &:=& \left. \frac{e^{\frac{i}{2} \prt_{1} \wedge \prt_{2}} - 1}{\frac{i}{2} \prt_{1} \wedge \prt_{2}} f(x_{1}) g(x_{2}) \right|_{x_{1}=x_{2}} \pc \nonumber\\
    f \star_{1}^{s} g &:=& \frac{1}{2} \{ f \stackrel{\star_{1}}{,} g \} = \left. \frac{\sin\left(\frac{1}{2} \prt_{1} \wedge \prt_{2}\right)}{\frac{1}{2} \prt_{1} \wedge \prt_{2}} f(x_{1}) g(x_{2}) \right|_{x_{1}=x_{2}} \pt
\eea Clearly, the terms tend to get more complicated at each order,
which makes higher order calculations via $\theta$-exact
Seiberg-Witten map highly elaborate.

Since $(f \star g)^{\dagger} = g^{\dagger} \star f^{\dagger}$ for
any functions (or, more generally, matrices) $f$ and $g$,
$\hat{\bar{\Psi}} \equiv \bar{\hat{\Psi}} =
\hat{\Psi}^{\dagger}\gamma^{0}$. Substituting (\ref{eq:hatA}) and
(\ref{eq:hatPsi}) into the action (\ref{eq:ncqedaction}), we find
the first order fermion-photon interaction term to be \bea
    \mathcal{L}_{\bar{\Psi} A \Psi}^{(1)} &=& -e\bar{\Psi} \star A\fsl \star \Psi \nonumber\\
    && - \frac{e}{2} \theta^{\alpha\beta} \left[ (\prt_{\beta}\bar{\Psi}) \star_{1} A_{\alpha} + \frac{1}{2} \bar{\Psi} \star_{1} (\prt_{\beta}A_{\alpha}) \right] \star (i\prt\fsl - m) \Psi \nonumber\\
    && - \frac{e}{2} \theta^{\alpha\beta} \bar{\Psi} \star (i\prt\fsl - m) \left[ A_{\alpha} \star_{1} (\prt_{\beta}\Psi) + \frac{1}{2} (\prt_{\beta}A_{\alpha}) \star_{1} \Psi \right] \pt
\eea Similarly, from (\ref{eq:hatA}) we find the photon-photon
interaction Lagrangian up to the first order in $e$ to be \bea
    \mathcal{L}_{A^{3}}^{(1)} &=& -\frac{e}{4} \Big\{ \prt_{\mu}A_{\nu} - \prt_{\nu}A_{\mu} \strc i\left[ A^{\mu} \strc A^{\nu} \right] \nonumber\\
    && \qquad -\frac{1}{2}\theta^{\alpha\beta} \Big[ \prt^{\mu} \Big( A_{\alpha} \star_{1}^{s} (2\prt_{\beta}A^{\nu} - \prt^{\nu}A_{\beta}) \Big) \nonumber\\
    && \qquad \qquad \quad - \prt^{\nu} \Big( A_{\alpha} \star_{1}^{s} (2\prt_{\beta}A^{\mu} - \prt^{\mu}A_{\beta}) \Big) \Big] \Big\} \pt
\eea The second order contributions are considerably more
complicated, but are obtained similarly by substituting the
expressions (\ref{eq:hatA}) and (\ref{eq:hatPsi}) into the action
(\ref{eq:ncqedaction}) and picking up the terms with the factor
$e^{2}$. The Feynman diagram vertex functions arising from the
first- and second-order parts of the action are given in Appendix
\ref{app}.

\section{Photon self-energy corrections}
Now, using the vertex function (\ref{eq:vertex}) from Appendix
\ref{app}, we find for the second order fermion loop correction to
the photon propagator arising from the diagram (a) in Fig.\!\!
\ref{fig:selfenergy} the form:
\bea
    \Pi_{\mt{(a)}}^{\alpha\beta}(k) &=& -4e^{2}\int \frac{\dd^{4}p}{(2\pi)^{4}} \nonumber\\
    && \times \Bigg\{ T^{\alpha\beta} + \frac{i}{2} \frac{\sin(\frac{1}{4} p \wedge k)}{\frac{1}{4} p \wedge k} \Bigg[ (\tilde{p} - \frac{1}{2}\tilde{k})^{\alpha} k_{\rho} T^{\rho\beta} e^{-\frac{i}{4} p \wedge k} \nonumber\\
    && \qquad \qquad \qquad - (\tilde{p} - \frac{1}{2}\tilde{k})^{\beta} k_{\rho} T^{\rho\alpha} e^{\frac{i}{4} p \wedge k} \Bigg] \nonumber\\
    && \qquad + \frac{1}{4} \frac{\sin^{2}(\frac{1}{4} p \wedge k)}{(\frac{1}{4} p \wedge k)^{2}} (\tilde{p} - \frac{1}{2}\tilde{k})^{\alpha} (\tilde{p} - \frac{1}{2}\tilde{k})^{\beta} k_{\rho} k_{\sigma} T^{\rho\sigma} \Bigg\} \pc \label{eq:propcorr}\nonumber\\
\eea where \be
    T^{\alpha\beta} := \frac{(p-k)^{\alpha} p^{\beta} + p^{\alpha} (p-k)^{\beta} + [ m^{2} - (p-k) \cdot p ] \eta^{\alpha\beta}}{p^{2} (p-k)^{2}} \pc
\ee which is the only term we get in the commutative case. Here we
see that we obtain the commutative result at the limit $\theta
\rightarrow 0$ in contrast with the result of Schupp and You
\cite{SchuppYou}. This follows from their use of the adjoint
representation for the spinor field, leading one to antisymmetrize
the $\star$-products in the action, which gives rise to a sine phase
factor for the vertex function. In our case of the fundamental
representation for the fermions, on the other hand, one does not
antisymmetrize the $\star$-products, thus obtaining an exponential
phase factor. The exponential factors cancel out upon the
multiplication of complex conjugates arising from the two vertices
of the diagram (a), leading to the usual commutative contribution,
while the sine factors arising from the vertices in the adjoint
representation do not cancel out upon multiplication. The first term
in (\ref{eq:propcorr}) along with parts of the second and the third
terms without phase factors can be interpreted as corresponding to
the planar contribution of the diagram. The parts of the second and
the third terms with nontrivial phase factors, on the other hand,
correspond to the nonplanar part, which typically gives rise to
UV/IR mixing.

To evaluate the parts of the second term in (\ref{eq:propcorr}) with
phase factors, we use the trick of Schupp and You \cite{SchuppYou}
by expressing them as
\bea
    && 2\int\!\!\frac{\dd^{4}p}{(2\pi)^{4}} \frac{1}{\frac{1}{2}p \wedge k}\left[ (\tilde{p} - \frac{1}{2}\tilde{k})^{\mu} k_{\rho} T^{\rho\nu} e^{-\frac{i}{2} p \wedge k} + (\tilde{p} - \frac{1}{2}\tilde{k})^{\nu} k_{\rho} T^{\rho\mu} e^{\frac{i}{2} p \wedge k} \right] \nonumber\\
    &=& 2i \sum_{\lambda= \pm 1} \int\!\!\dd \lambda\ I^{\mu\nu}(k;\lambda) \pc
\eea where \be
    I^{\mu\nu}(k;\lambda) = \int\!\frac{\dd^{4}p}{(2\pi)^{4}} ( \tilde{p} - \frac{1}{2} \tilde{k} )^{\mu} k_{\rho} T^{\rho\nu} e^{\frac{i}{2}\lambda p \wedge k} \pt
\ee By performing a Wick rotation $p^{\mu} = e^{\mu}_{i}
\bar{p}^{i}$, where $e^{\mu}_{i} = \mt{diag}(i,1,1,1)$ and
$\bar{p}^{i}$ is the Euclidean momentum, and using Schwinger
parametrization \be
    \frac{1}{\bar{p}^{2} + m^{2}} = \int_{0}^{\infty}\!\! \dd\alpha\ e^{-\alpha(\bar{p}^{2} + m^{2})} \pc
\ee we get \bea
    I^{\mu\nu}(k;\lambda) &=& ie^{\mu}_{i}e^{\nu}_{j} \iint_{0}^{\infty}\!\! \dd\alpha \dd\beta \int\!\frac{\dd^{4}p}{(2\pi)^{2}} (\bar{\tilde{p}} - \frac{1}{2}\bar{\tilde{k}})^{i} \left[ (\bar{k}^{2} - 2\bar{k} \cdot \bar{p}) \bar{p}^{j} + (\bar{p}^{2} + m^{2})\bar{k}^{j} \right] \nonumber\\
    && \qquad \qquad \times\ e^{-\alpha [(\bar{p} - \bar{k})^{2} + m^{2}] - \beta[\bar{p}^{2} + m^{2}] + \frac{i}{2} \lambda \bar{p} \cdot \bar{\tilde{k}}} \pt
\eea We may render the momentum integral Gaussian by applying the
change of variables \be
    \bar{q} := \bar{p} - \frac{\alpha}{\alpha + \beta} \bar{k} - \frac{i \lambda}{4(\alpha + \beta)} \bar{\tilde{k}} \pc
\ee after which we can perform the integration over $\bar{q}$.
Further, multiplying the integrand by \be
    1 = \int_{0}^{\infty} \!\!\dd c\ \delta(c-\alpha-\beta) \pc
\ee changing the order of integrations, and applying the change of
variables $\alpha = ca$, $\beta = cb$, we get \bea
    I^{\mu\nu}(k;\lambda) & \approx & \frac{ie_{i}^{\mu}e_{j}^{\nu} \bar{\theta}^{ik}}{(4\pi)^{2}} \iint_{0}^{1}\!\!\dd a \dd b\ \delta(1-a-b) \int_{0}^{\infty}\!\!\dd c\ c^{-3} \nonumber\\
    && \times \left[ \left( \frac{i\lambda}{2} - \frac{i\lambda^{3}\bar{\tilde{k}}^{2}}{64c}  \right) \bar{\tilde{k}}_{k} \bar{k}^{j} - \frac{i\lambda}{4} \bar{k}_{k}\bar{\tilde{k}}^{j} \right] \nonumber\\
    && \times\ e^{-c(ab\bar{k}^{2} + m^{2}) - \frac{\lambda^{2}}{16c}\bar{\tilde{k}}^{2}} \pc
\eea where the less IR-divergent terms are dropped
out.\footnote{Here we have to take into account the following
integration over $c$, where $c \sim \bar{\tilde{k}}^{2}$. The
integration over $\lambda$ does not affect the relative powers of
divergence.} The dependence on $a$ and $b$ drops out, and the
integrals over them give unity. The integral over $\lambda$ is now
straightforward to perform. Moreover, the integral over $c$ can be
performed and expressed for small $k$ using the properties of
modified Bessel functions $K_{r}(x,y)$ \cite{Magnus}: \bea
    \int_{0}^{\infty}\!\!\dd c\ c^{-r-1} e^{-xc-y/c} &=& 2\left(\frac{x}{y}\right)^{\frac{r}{2}} K_{r} \left[ 2\sqrt{xy} \right] \whr \mt{Re}[x],\mt{Re}[y] > 0 \pc \nonumber\\
    \mand K_{r}(z) & \approx & \frac{\Gamma(r)}{2}\left(\frac{2}{z}\right)^{r} \quad \mt{, when}\quad 0 < z \ll \sqrt{r+1} \pt
\eea We get for small $\bar{\tilde{k}}^{2} \ll m^{-2}$ accordingly
\be\label{eq:nppropcorr}
    i\Pi_{\mt{(a)np}}^{\mu\nu}(k) \approx \frac{8e^{2}}{\pi^{2}}\frac{\tilde{k}^{\mu} \tilde{k}^{\nu}}{\tilde{k}^{4}} + \frac{4e^{2}}{\pi^{2}}\frac{\tilde{\tilde{k}}^{\mu}k^{\nu} + k^{\mu}\tilde{\tilde{k}}^{\nu}}{\tilde{k}^{4}} \pt
\ee The first term here is similar to the IR-divergent terms found
in the usual formulation of NCQED and by Schupp and You
\cite{SchuppYou}. The second term gives another quadratic
IR-divergence, which is gauge variant, and therefore should be
cancelled, when all the second order contributions in the coupling
constant are taken into account.

Having found the gauge invariant IR-divergence in
(\ref{eq:nppropcorr}), we proceed to confirm the absence of
canceling terms. The calculations, though more elaborate, follow
precisely the same scheme as the one above. To make them manageable
we only consider contributions of the form $a \tilde{k}^{\mu}
\tilde{k}^{\nu}$, where $a$ is a scalar quantity. The third term in
(\ref{eq:propcorr}) and the other second order contributions coming
from the diagrams (b), (c) and (d) also give rise to quadratic
divergencies of the form $c \tilde{k}^{\mu} \tilde{k}^{\nu} /
\tilde{k}^{4}$, where $c$ is a constant. For all of the
contributions we find $c > 0$, and thus they cannot cancel the
IR-divergence of (\ref{eq:nppropcorr}). Hence we conclude that the
UV/IR mixing problem persists in the noncommutative QED formulated
here via $\theta$-exact Seiberg-Witten map.

\section{Conclusions and remarks}
We have found that UV/IR mixing is present in the photon self-energy
corrections of noncommutative QED defined via $\theta$-exact
Seiberg-Witten map for a straight path in $\theta$-space. The result
further demonstrates that UV/IR mixing is a generic property of
noncommutative quantum field theories, and is not cured in general
by the approach via Seiberg-Witten map, contrary to some claims
previously made in the literature.

A question remains open, though, whether the result holds generally
for all possible integration paths in $\theta$-space. It is not
ruled out that by modifying the integration path one could get rid
of the divergence, although on mathematical grounds this seems
unlikely, at least, for paths obtainable from the straight one by
smooth deformations. Of course, answering the question properly
requires a rigorous analysis, which we postpone to a future study.

In the case of a scalar field theory in noncommutative spacetime, as
for example in \cite{Minwalla}, the destruction of UV/IR mixing by
$\theta$-expansion becomes immediately obvious. In \cite{Minwalla}
there is also given a satisfactory explanation for the mixing as a
direct result of the infinite nonlocality of $\star$-product, and
thus it is deeply rooted in the very definition of noncommutativity
of spacetime. It therefore seems unlikely that the resulting
divergencies could be made vanish, at least, without modifying the
theory itself by introducing new terms in the Lagrangian, which
suppress the contributions of the IR sector. This has been done for
a noncommutative scalar field theory in \cite{Grosse1,Grosse2,Gurau}
and for noncommutative QED in \cite{BlaschkeRofner,Vilar}. For some
other attempts, see \cite{Fisher,DenkSchweda,DenkPutz}. Reducing the
nonlocality of the noncommutative field theories to a finite range
is also an option which has been preliminarily exploited in
\cite{BW,LZ}.

All in all, what ever the direction, more work is to be done before
we are to overcome the obstacles arising from the nonlocality in
noncommutative quantum field theories.

\section*{Acknowledgements}
We would like to thank Masud Chaichian for many helpful discussions.
The support of the Academy of Finland under the Projects No. 121720
and 127626 is gratefully acknowledged.

\appendix

\section{Vertex functions}\label{app}
The momenta are everywhere incoming, $p_{i}$'s for fermions and
$k_{i}$'s for photons.

For the first order vertex functions we get the expressions: \be
\label{eq:vertex}
    V_{\bar{\Psi} A \Psi}^{\mu}(p_{1},p_{2}) = -ie\gamma^{\mu} e^{\frac{i}{2} p_{1} \wedge p_{2}} - \frac{ie}{2} (\tilde{p}_{1} - \tilde{p}_{2})^{\mu} (p_{1}\!\!\fsl\; + p_{2}\!\!\fsl) \frac{e^{\frac{i}{2}p_{1} \wedge p_{2}} - 1}{p_{1} \wedge p_{2}} \pc
\ee \bea
    && V_{A^{3}}^{\mu_{1}\mu_{2}\mu_{3}}(k_{1},k_{2},k_{3}) \nonumber\\
    &=& 2e \sin\left(\frac{k_{1} \wedge k_{2}}{2} \right) \nonumber\\
    && \quad \times \Big\{ (k_{1} - k_{2})^{\mu_{3}} \eta^{\mu_{1}\mu_{2}} + \frac{1}{\frac{1}{2} k_{1} \wedge k_{2}} \Big[ (k_{1}^{\mu_{1}} k_{1}^{\alpha} - k_{1}^{2} \eta^{\mu_{1}\alpha}) (2\tilde{k}_{3}^{\mu_{2}} \eta_{\alpha}^{\mu_{3}} - k_{3\alpha} \theta^{\mu_{2}\mu_{3}}) \Big] \Big\} \nonumber\\
    && +\ \{symm.\} \pc
\eea where $\{symm.\}$ denotes terms symmetrizing the previous
contributions with respect to the photons $(k_{i},\mu_{i})$.

For the second order vertex functions we similarly find \bea
    && V^{\mu_{1}\mu_{2}}_{\bar{\Psi} A^{2} \Psi}(k_{1},k_{2},p_{1},p_{2}) \nonumber\\
    &=& -\frac{ie^{2}}{4} \Bigg\{ \Bigg[ \Bigg( \frac{\sin(\frac{1}{2} k_{1} \wedge k_{2})}{\frac{1}{2} k_{1} \wedge k_{2}} \frac{e^{\frac{i}{2}p_{1} \wedge p_{2}}}{\frac{i}{2}p_{1} \wedge p_{2}} - \frac{\cos(\frac{1}{2}k_{1}\wedge k_{2}) e^{\frac{i}{2}p_{1} \wedge p_{2}} - 1}{(\frac{i}{2}p_{1} \wedge p_{2})^{2}} \Bigg) \nonumber\\
    && \qquad \qquad \times \Big( (\tilde{p}_{1} - \tilde{p}_{2})^{\mu_{1}} \tilde{k}_{1}^{\mu_{2}} - \frac{1}{2} \theta^{\mu_{1}\mu_{2}} (p_{1} - p_{2}) \wedge k_{1} \Big) \nonumber\\
    && \qquad \quad + \Bigg( \frac{e^{\frac{i}{2}p_{2} \wedge k_{1}}}{\frac{i}{2}p_{2} \wedge k_{1}} \frac{e^{\frac{i}{2}k_{2} \wedge p_{1}} - 1}{\frac{i}{2}k_{2} \wedge p_{1}} - \frac{e^{\frac{i}{2}p_{2} \wedge k_{1}} e^{\frac{i}{2}k_{2} \wedge p_{1}} - 1}{(\frac{i}{2}p_{2} \wedge k_{1})^{2}} \Bigg) \nonumber\\
    && \qquad \qquad \times \frac{1}{4}(\tilde{p}_{1} - \tilde{p}_{2} + \tilde{k}_{2})^{\mu_{1}} (\tilde{p}_{1} - \tilde{p}_{2} - \tilde{k}_{1})^{\mu_{2}} \Bigg] (p_{2}\!\!\fsl\ - m) \nonumber\\
    && \qquad - \Bigg[ \Bigg( \frac{\sin(\frac{1}{2} k_{1} \wedge k_{2})}{\frac{1}{2} k_{1} \wedge k_{2}} \frac{e^{\frac{i}{2}p_{1} \wedge p_{2}}}{\frac{i}{2}p_{1} \wedge p_{2}} - \frac{\cos(\frac{1}{2}k_{1}\wedge k_{2}) e^{\frac{i}{2}p_{1} \wedge p_{2}} - 1}{(\frac{i}{2}p_{1} \wedge p_{2})^{2}} \Bigg) \nonumber\\
    && \qquad \qquad \times \Big( (\tilde{p}_{2} - \tilde{p}_{1})^{\mu_{2}} \tilde{k}_{2}^{\mu_{1}} - \frac{1}{2} \theta^{\mu_{2}\mu_{1}} (p_{2} - p_{1}) \wedge k_{2} \Big) \nonumber\\
    && \qquad \quad + \Bigg( \frac{e^{\frac{i}{2}p_{2} \wedge k_{1}}}{\frac{i}{2}p_{2} \wedge k_{1}} \frac{e^{\frac{i}{2}k_{2} \wedge p_{1}} - 1}{\frac{i}{2}k_{2} \wedge p_{1}} - \frac{e^{\frac{i}{2}p_{2} \wedge k_{1}} e^{\frac{i}{2}k_{2} \wedge p_{1}} - 1}{(\frac{i}{2}p_{2} \wedge k_{1})^{2}} \Bigg) \nonumber\\
    && \qquad \qquad \times \frac{1}{4}(\tilde{p}_{2} - \tilde{p}_{1} + \tilde{k}_{1})^{\mu_{2}} (\tilde{p}_{2} - \tilde{p}_{1} - \tilde{k}_{2})^{\mu_{1}} \Bigg] (p_{1}\!\!\fsl\ + m) \nonumber\\
    && \qquad + \frac{1}{4} \frac{e^{\frac{i}{2}k_{1}\wedge p_{1}} - 1}{\frac{i}{2}k_{1}\wedge p_{1}} \frac{e^{\frac{i}{2}k_{2}\wedge p_{2}} - 1}{\frac{i}{2}k_{2}\wedge p_{2}} (\tilde{p}_{1} - \tilde{p}_{2} - \tilde{k}_{2})^{\mu_{1}} (\tilde{p}_{2} - \tilde{p}_{1} - \tilde{k}_{1})^{\mu_{2}} (p_{1}\!\!\fsl\ + k_{1}\!\!\fsl\ + m) \nonumber\\
    && \qquad + i\frac{e^{\frac{i}{2}k_{1}\wedge p_{1}} - 1}{\frac{i}{2}k_{1}\wedge p_{1}} e^{\frac{i}{2}p_{2}\wedge k_{2}} (\tilde{p}_{1} - \tilde{p}_{2} - \tilde{k}_{2})^{\mu_{1}} \gamma^{\mu_{2}} \nonumber\\
    && \qquad + i\frac{e^{\frac{i}{2}k_{2}\wedge p_{2}} - 1}{\frac{i}{2}k_{2}\wedge p_{2}} e^{\frac{i}{2}p_{1}\wedge k_{1}} (\tilde{p}_{2} - \tilde{p}_{1} - \tilde{k}_{1})^{\mu_{2}} \gamma^{\mu_{1}} \nonumber\\
    && \qquad + 4i \frac{\sin(\frac{1}{2}k_{1} \wedge k_{2})}{\frac{1}{2}k_{1} \wedge k_{2}} e^{\frac{i}{2} p_{1} \wedge p_{2}} (\tilde{k}_{2}^{\mu_{1}} \gamma^{\mu_{2}} - \frac{1}{2}\theta^{\mu_{1}\mu_{2}} k_{2}\!\!\fsl\ ) \Bigg\} \nonumber\\
    && +\ \{symm.\} \pc
\eea

\bea
    && V^{\mu_{1}\mu_{2}\mu_{3}\mu_{4}}_{A^4}(k_{1},k_{2},k_{3},k_{4}) \nonumber\\
    &=& -e^{2} \Bigg\{ \Bigg[ \frac{\sin(\frac{1}{2}k_{1} \wedge k_{2})}{\frac{1}{2}k_{1} \wedge k_{2}} \frac{\sin(\frac{1}{2}k_{3} \wedge k_{4})}{\frac{1}{2}k_{3} \wedge k_{4}} - \frac{\cos(\frac{1}{2}k_{1} \wedge k_{2}) \cos(\frac{1}{2}k_{3} \wedge k_{4}) - 1}{(\frac{1}{2}k_{3} \wedge k_{4})^{2}} \Bigg] \nonumber\\
    && \qquad \quad \times \Big[ \Big( \tilde{k}_{2}^{\mu_{1}} \tilde{k}_{3}^{\mu_{2}} - \frac{1}{2}\theta^{\mu_{1}\mu_{2}} k_{2} \wedge k_{3} \Big) \eta^{\mu_{3}}_{\alpha} \nonumber\\
    && \qquad \qquad + \Big( (\tilde{k}_{1} + \tilde{k}_{2})^{\mu_{3}} \tilde{k}_{2}^{\mu_{1}} - \frac{1}{2}\theta^{\mu_{3}\mu_{1}} k_{1} \wedge k_{2} \Big) \eta^{\mu_{2}}_{\alpha} \nonumber\\
    && \qquad \qquad + \frac{1}{2} \Big( \theta^{\mu_{2}\mu_{3}} \tilde{k}_{2}^{\mu_{1}} + \theta^{\mu_{1}\mu_{3}} \tilde{k}_{1}^{\mu_{2}} - \frac{1}{2} \theta^{\mu_{1}\mu_{2}}(3\tilde{k}_{1} + \tilde{k}_{2})^{\mu_{3}} \Big) k_{2\alpha} \nonumber\\
    && \qquad \qquad - \frac{1}{2} \Big( \theta^{\mu_{2}\mu_{3}}\tilde{k}_{2}^{\mu_{1}} + \frac{1}{2} \theta^{\mu_{1}\mu_{2}}\tilde{k}_{2}^{\mu_{3}} \Big) k_{3\alpha} \Big] \nonumber\\
    && \qquad \quad \times \Big(k_{4}^{2}\eta^{\alpha\mu_{4}} - k_{4}^{\alpha} k_{4}^{\mu_{4}} \Big) \nonumber
\eea \bea
    && \qquad - \frac{1}{2} \frac{\sin(\frac{1}{2}k_{1} \wedge k_{2})}{\frac{1}{2}k_{1} \wedge k_{2}} \frac{\sin(\frac{1}{2}k_{3} \wedge k_{4})}{\frac{1}{2}k_{3} \wedge k_{4}} \Big( \eta^{\mu_{2}}_{\alpha} \tilde{k}_{2}^{\mu_{1}} - \frac{1}{2}\theta^{\mu_{1}\mu_{2}} k_{2\alpha} \Big) \nonumber\\
    && \qquad \quad \times \Big( \eta^{\mu_{4}}_{\beta} \tilde{k}_{4}^{\mu_{3}} - \frac{1}{2}\theta^{\mu_{3}\mu_{4}} k_{4\beta} \Big) \Big( (k_{1} + k_{2})^{2} \eta^{\alpha\beta} - (k_{1} + k_{2})^{\alpha} (k_{1} + k_{2})^{\beta} \Big) \nonumber\\
    && \qquad + \frac{\sin(\frac{1}{2}k_{1} \wedge k_{2})}{\frac{1}{2}k_{1} \wedge k_{2}} \sin(\frac{1}{2}k_{3} \wedge k_{4}) \Big( \eta^{\mu_{2}}_{\alpha} \tilde{k}_{2}^{\mu_{1}} - \frac{1}{2}\theta^{\mu_{1}\mu_{2}} k_{2\alpha} \Big) \nonumber\\
    && \qquad \quad \times \Big( \eta^{\mu_{3}\mu_{4}} k_{4}^{\alpha} - \eta^{\alpha\mu_{4}} k_{4}^{\mu_{3}} \Big) \nonumber\\
    && \qquad + \sin(\frac{1}{2}k_{1} \wedge k_{2}) \frac{\sin(\frac{1}{2}k_{3} \wedge k_{4})}{\frac{1}{2}k_{3} \wedge k_{4}} \Big( \eta^{\mu_{4}}_{\alpha} \tilde{k}_{4}^{\mu_{3}} - \frac{1}{2}\theta^{\mu_{3}\mu_{4}}k_{4\alpha} \Big) \nonumber\\
    && \qquad \quad \times \Big( \eta^{\mu_{2}}_{\alpha} (k_{3} + k_{4})^{\mu_{1}} - \eta^{\mu_{1}}_{\alpha} (k_{3} + k_{4})^{\mu_{2}} \Big) \nonumber\\
    && \qquad + \sin(\frac{1}{2}k_{1} \wedge k_{2}) \sin(\frac{1}{2}k_{3} \wedge k_{4}) \eta^{\mu_{1}\mu_{2}} \eta^{\mu_{3}\mu_{4}} \Bigg\} \nonumber\\
    && +\ \{symm.\} \pc
\eea where again $\{symm.\}$ denotes terms symmetrizing the previous
contributions with respect to the photons $(k_{i},\mu_{i})$.

\end{document}